# A DYNAMIC MST- deltaCoVar MODEL OF SYSTEMIC RISK IN THE EUROPEAN INSURANCE SECTOR.

Anna Denkowska and Stanisław Wanat

## Abstract

This work is an answer to the EIOPA 2017 report. It follows from the latter that in order to assess the potential systemic risk we should take into account the build-up of risk and in particular the risk that arises in time, as well as the interlinkages in the financial sector and the whole economy.
Our main tools used to analyse the systemic risk dynamics in the European insurance sector during the years 2005-2019 are the topological indices of minimum spanning trees (MST) and the deltaCoVaR measure.
We address the following questions:
1) What is the contribution to systemic risk of each of the 28 largest European insurance companies whose list includes also those appearing on the G-SIIs list?
2) Does the analysis of the deltaCoVaR of those 28 insurance companies and the conclusions we draw agree with the our claims from our latest article [Wanat S., Denkowska A. 2019]. In clear: does the most important contribution to systemic risk come from the companies that have the highest betweenness centrality or the highest degree in the MST obtained?

## Introduction

The subject of this work is part of the current research on the interlinkages of large insurance companies and their contribution to systemic risk in the insurance sector. In international literature we find many studies related to the systemic risk analysis in the banking sector but little is done regarding systemic risk in the insurance sector. A novelty we bring to the literature is to identify the relationship between the contribution to systemic risk and the structure of the minimum spanning tree described by topological network indicators, which can be used in further research to construct models whose task is to predict the possibility of systemic risk.

Our work constitutes an answer to the recommendations contained in the 2017 report of the European Insurance and Occupational Pensions Authority (EIOPA), an independent EU advisory body to the European Parliament, the Council of Europe and the European Commission, which shows that when analyzing systemic risk in the insurance sector, one should take into account, among others, the dynamics of interconnectedness between institutions. The present article is another study of the authors in this subject. In a previous paper [Wanat, Denkowska (2019)], selected topological indicators of Minimum Spanning Trees were analysed and interrelationships between major European insurance institutions were examined.

The purpose of this work is to confront the contribution to the systemic risk of each of the 28 insurance institutions analysed with the results of a previous study. We examine whether interlinkages between insurers, and thus the level of possibility of contagion with a potential crisis, is related to the creation of systemic risk by individual insurance companies.

The global economic crisis on financial markets, the peak of which was in the years 2008-2009, initiated on the market of high-risk mortgage loans in the USA as a result, among others, of the deregulation of the financial market in 1999, when the ban on combining two types of banking: investment (high risk) and deposit-credit banking, which was in force since 1929 and was to protect citizens in the event of losses in investment banking, ceased to apply. On September 15th 2008 Lehman Brothers' – the fourth largest investment bank, after a fruitless attempt to obtain help from the US central bank, declared bankruptcy. A week earlier, the Fed took over two insurance and loan companies with huge debts: Fannie Mae and Freddie Mac. The Fed and the Ministry of the Treasury have recapitalized the largest insurance and financial holding company AIG in the amount of USD 85bn, as the loss of liquidity and, consequently, its collapse would mean a rapid spread of the crisis. The American International Group (AIG) is the first example of an insurance company that required (and received) funding because it was considered systemically important. When dividing financial institutions into three groups according to their activity, i.e. investment, depository and risk-dissipating institutions, insurance companies should be in the third group. Therefore, they should not generate systemic risk as long as they deal with taking over, dispersing and redistributing the financial effects of risk. But if they take over credit risk, e.g. financial insurance, insurance guarantees, derivative trading, in particular like AIG Credit Default swap (CDS) involved in trading financial instruments given as collateral in case of default on repayment obligations, they generate systemic risk (see [William K. Sjostrum, Jr. (2009)])

After the subprime crisis, all financial supervising authorities drew attention to the need for macro-prudential policy, which would take into account the dynamics of structure changes and linkages between financial institutions. The situation of AIG in 2008 was surprising, when it was announced in February that in 2007 it achieved profits of USD 6.20 billion (USD 2.39 per share). The stock was closed on that day at USD 50.15 per share. Less than seven months later, the company was on the verge of bankruptcy and had to be funded by the US government. This is the first and very important event that indicates that the insurance sector plays a large role when assessing systemic risk. Therefore, in 2013 IAIS, when developing a method of identifying insurance institutions of particular importance for financial stability, takes into account the following five dimensions (see [Global Systemically ...., 2013]):

• size of the insurance institution (5%),
• range of global activity (5%),
• assessment of the degree of direct and indirect linkages of the institution within the financial system (40%),
• non-traditional and non-insurance activity of the insurer (45%),
• product substitutability - the significance of the institution for the financial system increases along with the lack of real substitution possibilities for the services provided by the insurer (5%).

In line with the IAIS recommendations, the Financial Stability Board (FSB) announced in 2016 a list of systemically important insurers (G-SIIs): Aegon N.V., Allianz SE, American International Group, Inc. (AIG), Aviva plc, Axa S.A., MetLife, Inc., Ping An Insurance (Group) Company of China, Ltd., Prudential Financial, Inc., Prudential plc.

**Literature review**

Before the subprime (2007-2009) and excessive public debt in the euro area countries (2010-2013) crises, there was a strong belief that the insurance market is systemically insignificant. In international literature that emerged as a consequence of the crisis, many studies have maintained their previous beliefs, but we also find numerous works confirming the possibility of the insurance sector creating systemic risk. Examples include works in which authors believe that insurance companies have become an unavoidable source of systemic risk (e.g. [Billio et al. 2012], [Weiß, Mühlnickel 2014]) and those in which they claim that they can be systematically significant , but this is due to their non-traditional (banking) activities (e.g. [Baluch et al. 2011], [Cummins, Weiss 2014]) and the overall systemic importance of the insurance sector as a whole is still subordinated to the banking sector ([Chen et al. 2013] ). In turn, in [Bierth in. 2015], the authors, after examining a very large sample of insurers in the long term, believe that the contribution of the insurance sector to systemic risk is relatively small, however, they claim that it reached its peak during the financial crisis in 2007-2008, which we also confirm in our analysis. Analysts also report that significant factors affecting the insurers' exposure to systemic risk are the strong linkages of large insurance companies, leverage, losses and liquidity (four L's). However, the complete lack of evidence of the systemic importance of the insurance industry is indicated, among others, by the following papers [Harrington 2009], [Bell, Keller 2009] and [Geneva Association 2010].

The problem of the insurance sector's ability to create systemic risk has also been the subject of consideration in Polish scientific literature in recent years. In general, as above, two main positions are represented. In [Czerwińska 2014], based on research of the insurance sector in European Union countries covering the period 2005-2012, it was found that along with the increase in the level of linkages between insurers and various financial system segments, mainly the capital market and the banking sector, the importance of insurance institutions for the stability of the entire system increases. In turn, in [Bedarczyk 2013], the author, assessing insurance institutions as a potential creator of systemic risk, concluded that dispersing and taking over insurance risk ultimately does not create systemic risk. It indicates a relatively low level of interconnectedness and draws attention to the fact that insurers are not highly dependent on external financing, so they should not be included in the group of systemically important institutions. At the same time, it was mentioned that insurers engaged in non-insurance activities pose a threat to the system by taking over credit risk. CoVaR (see e.g. [Acharya et al. 2010], [Bierth et al. 2015], [Jobst 2014]). Measures of the impact (contribution) of an individual financial institution on the systemic risk of a given market and measures of the institution's sensitivity to this risk are CoVaR and ΔCoVaR [Adrian, Brunnermeier 2011]

## Empirical strategy

In order to identify the possible relationship linking the structure of the interconnections between insurance companies to the creation of systemic risk by these companies, we proceed in two steps. In the first one, we analyse the dynamics of the structure of interlinkages between insurers. For this purpose, we use the time series of the following selected topological network indicators (see Wang et al., 2014):
- Average Path Length (APL)
- Maximum Degree - Max.Degree
- parameters $\alpha$ of the power distribution of vertex degrees: $P(s) = C \cdot s^{-\alpha}$, $\alpha > 0$,
- betweenness centrality (BC).

The Average Path Length (APL) is defined as the average number of steps taken along all the shortest paths connecting all possible pairs of network nodes.

The Maximum Degree: in graph theory it is defined as the maximal number of edges coming out from a vertex (where each loop counts for two). In other words, it measures the number of connections to the central vertex.

The Betweenness Centrality (BC) measures the centrality of a vertex: we consider the ratio between the number of shortest paths connecting two vertices and passing through the given one, and the number of all the shortest paths between pairs of distinct vertices. It indicates thus the most important nodes of a network based on shortest paths (e.g. the most influential insurer).

We obtain these series based on the determined minimum spanning trees $MST_t$ for each period studied. We construct the $MST_t$ trees using conditional correlations between each pair of analysed insurance companies determined using the copula-DCC-GARCH model. Details of the construction are presented in the work [Wanat S., Denkowska A.2019].

In the second step, we examine the contribution of a single insurer to the systemic risk of the European insurance sector using the CoVaR delta model. In this model, the basis for measuring risk is the CoVaR measure. Formally, $CoVaR^{j|i}_{\beta,t}$ is defined as the value at risk (VaR) of an institution j under the condition that another institution i is at risk of crisis in a given period t, i.e. its rate of return is less than its value at risk:

$$P(r_{j,t} \leq CoVaR^{j|i}_{\beta,t} | r_{i,t} \leq VaR^{i}_{\alpha,t}) = \beta \tag{1}$$

Using the formula for the conditional probability we have:

$$\frac{P(r_{j,t} \leq CoVaR^{j|i}_{\beta,t}, r_{i,t} \leq VaR^{i}_{\alpha,t})}{P(r_{i,t} \leq VaR^{i}_{\alpha,t})} = \beta \tag{2}$$

In addition, the definition of value-at-risk for institutions and implies that

$$P(r_{i,t} \leq VaR^{i}_{\alpha,t}) = \alpha \tag{3}$$

From equations (2) and (3) we get:

$$P(r_{j,t} \leq CoVaR^{j|i}_{\beta,t}, r_{i,t} \leq VaR^{i}_{\alpha,t}) = \alpha\beta \tag{4}$$

From the relationship (4) we can estimate $CoVaR^{j|i}_{\beta,t}$, but first we need to determine the two-dimensional distribution $F_t$ of the rate of return vector $(r_{j,t}, r_{i,t})$. This distribution can be represented using the copula in the following ways:

$$F_t(r_{j,t}, r_{i,t}) = C_t(F(r_{j,t}), F(r_{i,t})) \tag{5}$$

From formula (5) we can numerically calculate $CoVaR^{j|i}_{\beta,t}$ by solving the equation:

$$C_t(F_t(CoVaR^{j|i}_{\beta,t}), \alpha) = \alpha\beta \tag{6}$$

Then, knowing the value of the measure $CoVaR^{j|i}_{\beta,t}$, we can calculate the measure deltaCoVaR ($\Delta CoVaR_\beta^{j|i}$). The value of this measure is the difference between the value at risk of the insurance sector (institution j), provided that the insurer (institution i) is in a state of financial crisis and the value at risk of the insurance sector in the event that the financial standing of the entity i is normal (average), i.e.

$$\Delta CoVaR_\beta^{j|i} = CoVaR_\beta^{j|X^i \leq VaR^i_\alpha} - CoVaR_\beta^{j|X^i \leq Median^i} \tag{7}$$

The value of this measure represents the contribution of the institution i to systemic risk. The lower this value, the greater the institution's share in generating systemic risk.

In our analysis we estimate the distributions $F_t$ of the vectors $(r_{j,t}, r_{i,t})$ and determine $CoVaR^{i|j}_{\beta,t}$ using the two-dimensional copula-DCC-GARCH models with the t-Student copula.

In these models, the average rate of return was modeled using the following ARIMA process:

$$r_{i,t} = \mu_{i,t} + y_{i,t} \tag{8}$$

$$\mu_{i,t} = E(r_{i,t}|\Omega_{t-1}) \tag{9}$$

$$\mu_{i,t} = \mu_{i,0} + \sum_{j=1}^{p} \varphi_{ij} r_{i,t-j} + \sum_{j=1}^{q} \theta_{ij} y_{i,t-j} \tag{10}$$

$$y_{i,t} = \sqrt{h_{i,t}} \varepsilon_{i,t}, \tag{11}$$

Where $\Omega_{t-1}$ denotes the collection of information available until the moment $t-1$, while $\varepsilon_{i,t}$ are independent random variables with identical distributions. We model the conditional variance h_ (i, t) using the exponential GARCH (eGARCH) model:

$$\log(h_{i,t}) = \omega_i + \sum_{j=1}^{p} \left( \alpha_{ij} \varepsilon_{i,t-j} + \gamma_{ij}(|\varepsilon_{i,t-j}| - E|\varepsilon_{i,t-j}|) \right) + \sum_{j=1}^{q} \beta_{ij} \log(h_{i,t-j}) \tag{12}$$

where $\varepsilon_{i,t} = \frac{y_{i,t}}{\sqrt{h_{i,t}}}$ stands for the standardized rests.

To model the relationship between the rates of return we use Student's t-copula, whose parameters are the conditional correlations $R_t$, obtained using the DCC model (m, n):

$$H_t = D_t R_t D_t \tag{13}$$

$$D_t = diag(\sqrt{h_{1,t}}, \dots, \sqrt{h_{k,t}}) \tag{14}$$

$$R_t = (diag(Q_t))^{-\frac{1}{2}} Q_t (diag(Q_t))^{-\frac{1}{2}} \tag{15}$$

$$Q_t = (1 - \sum_{j=1}^{m} c_j - \sum_{j=1}^{n} d_j)\bar{Q} + \sum_{k=1}^{m} c_j (\varepsilon_{t-j} \varepsilon'_{t-j}) + \sum_{k=1}^{n} d_j Q_{t-j} \tag{16}$$

$\bar{Q}$ is the unconditional covariance matrix for standardized remainders $\varepsilon_t$, $c_j$, $d_j$, $j = 1, \dots, m$ are scalar values, with $c_j$ describing the impact on current correlations of earlier shocks, and $d_j$ takes into account the impact on current correlations of earlier conditional correlations.

We estimate the parameters of the above copula-DCC-GARCH model using the inference function for margins (IFM) method.

## Data and results of empirical analysis

The basis of the analysis are stock quotes of 28 European insurance institutions selected from among the 50 largest[1]. Five of them, AXA, Allianz, Prudential plc, Aviva, Aegon appear as systemically relevant on the current list of G-SIIs published by the FSB in 2016. We analyse weekly logarithmic rates of return from January 7th, 2005 to April 26th, 2019.

Time series of topological network indicators, determined according to the first stage of the presented empirical strategy, are presented in Fig. 1 and 2. The first figure (Fig. 1) shows the average path length (APL), average maximum degree (Max.Degree) and estimated parameters $\alpha$ of the power distribution for minimum spanning trees from 07/01/2005 to 26/04/2019. The analysis of the charts shows that in the periods of June 2nd, 2006 - August 17th, 2007 and December 5th, 2008 - September 17th, 2010, APL decreases while Max.Degree increases, the α index is close to 2, which means that the network is shrinking and its structure is "scale-free", that is, it takes a form in which there are few vertices with numerous edges (hubs) and many vertices with a low degree (betweenness centrality). In turn, Figure 2 shows BC for AXA, Allianz and Phoenix. The first two companies have the highest average values of this measure in the examined period (see Fig. 3), while the third one (Phoenix) is one of the five insurers for which BC is equal to zero in each of the examined weeks. Considering the fact that BC is an indicator on the basis of which we assess the importance of a given insurer in the context of the possibility of risk contagion, we note that the time series for these companies are "complementary" behaving graphs. During the entire analysed period, if BC for AXA increases, BC for Allianz decreases and vice versa. If BC for AXA remains stable, the BC level for Allianz does not change either. Figure 3 shows that in the subprime crisis state the French AXA company was clearly the dominant one on the European insurance market, while the German Allianz took over during the phase of excessive public debt.

---

[1] We have chosen from among the 50 largest companies according to https://www.relbanks.com/top-insurance-companies/europe all those that were listed in the period studied, namely: **AXA**, **Allianz**, **Prudential plc**, Legal & General, Generali, **Aviva**, **Aegon**, CNP Assurances, Zurich Insurance, Munich Re, Old Mutual, Swiss Life, Chubb Ltd, Ageas, Phoenix, Unipol Gruppo, Mapfre, Hannover Re, Storebrand, XL.Group, Helvetia Holding, Vienna Insurance, SCOR SE, Mediolanum, Sampo Oyj, RSA Insurance Group, Società Cattolica di Assicurazione, Topdanmark A/S.

Fig.1 . Average distances (APL), maximum degrees (Max.Degree) and estimated parameters $\alpha$ of power distribution α for minimum spanning trees in the period 07.01.2005 - 26.04.2019

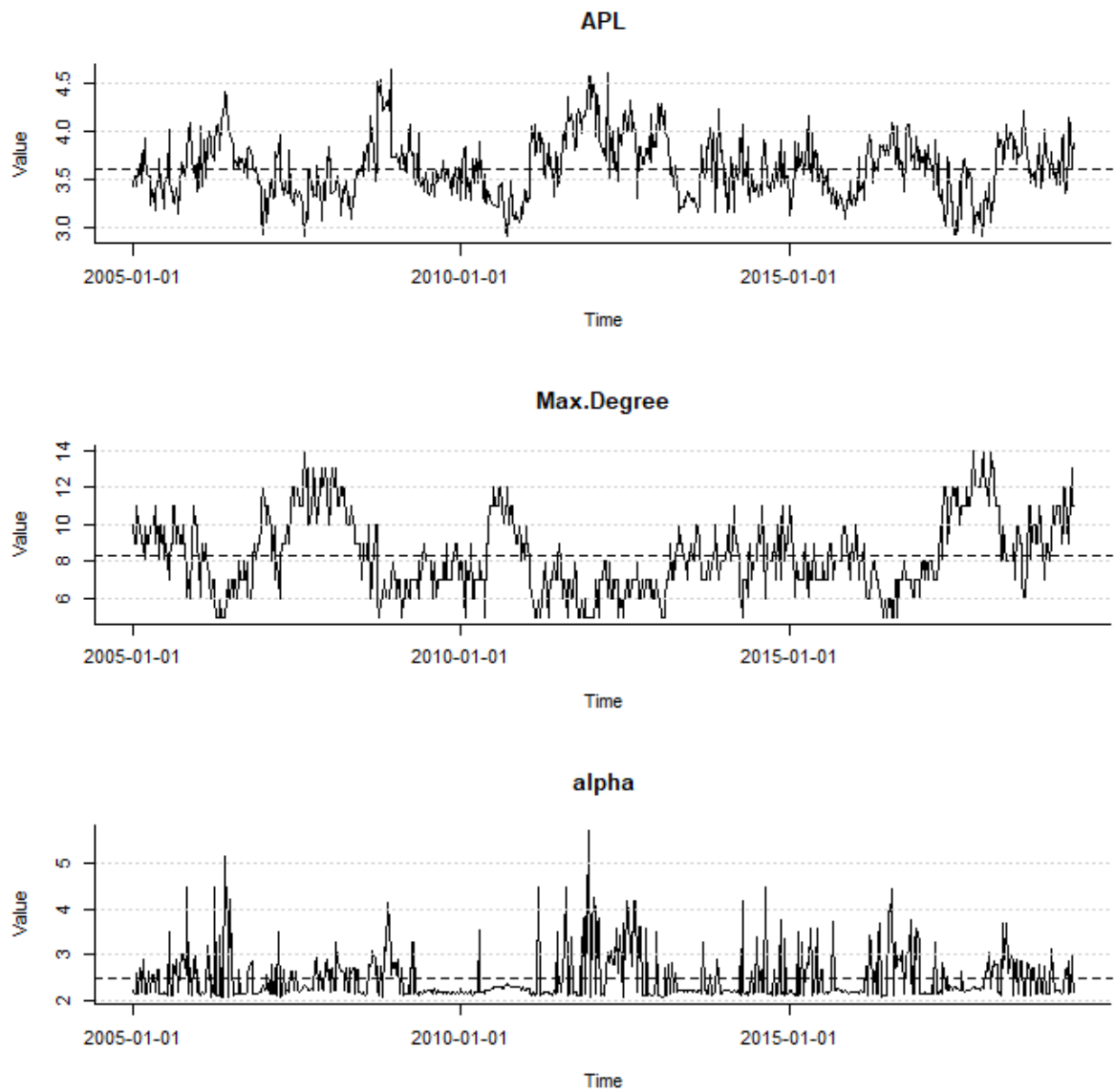

Source: Own study

Fig. 2. BC for selected insurance institutions (AXA, Allianz and Phoenix) during the period 07.01.2005 - 26.04.2019

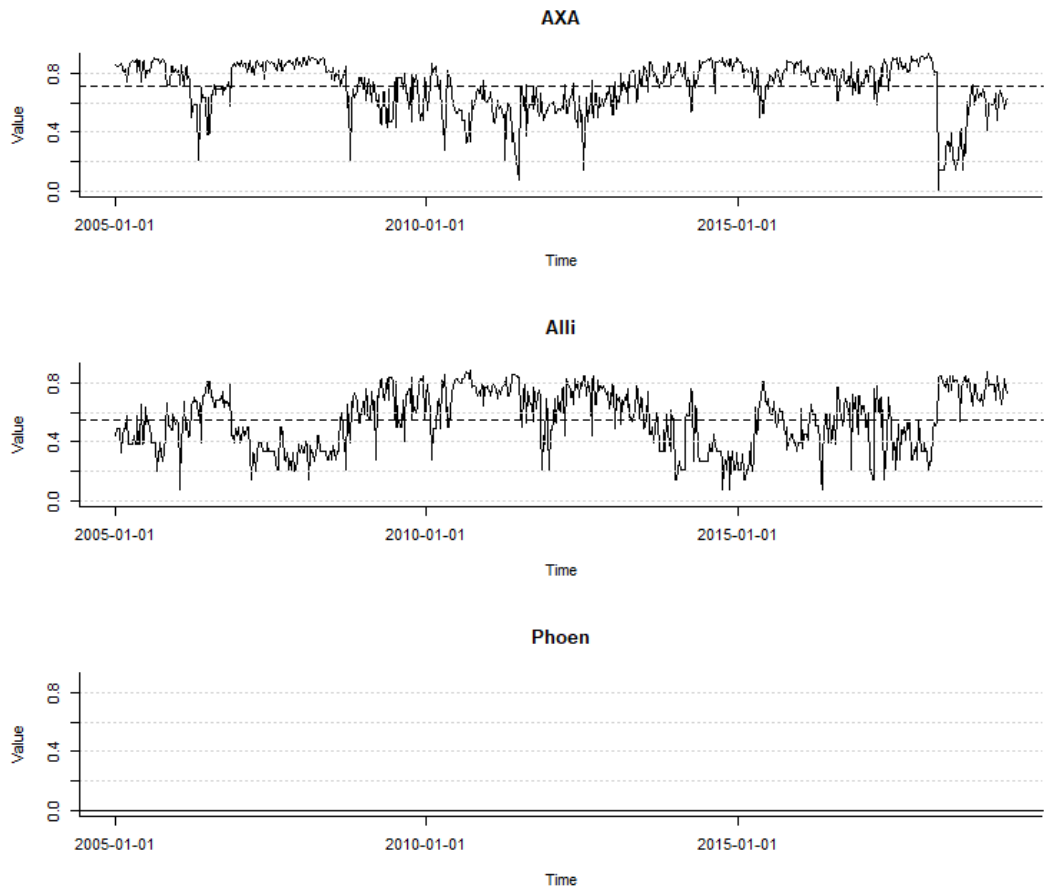

Source: Own study

Fig. 3  BC distribution for AXA and Allianz in the normal state of the market (N), during the subprime mortgage crisis (SMC) and the public debt crisis (PDC)

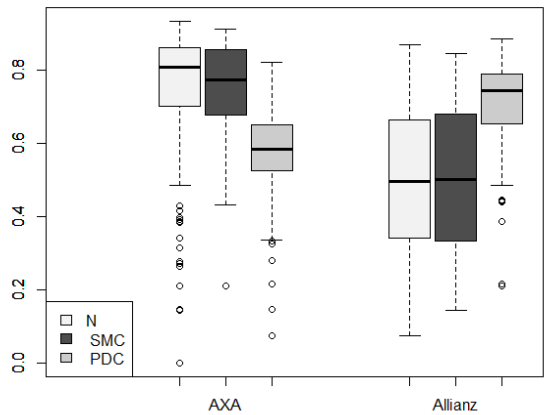

Source: Own study

We examine the relationship between the structure of the network of connections (MST) and the contribution to systemic risk based on time series deltaCoVaR measures, determined for individual insurers in accordance with the empirical strategy of the second step of the study. In each analysed period, we determine the deltaCoVaR measure for each insurer, assuming $\alpha = \beta = 0.05$. Figure 4 shows the average BC and Figure 5 shows the average deltaCoVaR over the period under consideration for all 28 companies analysed. Comparison of these diagrams shows that BC and deltaCoVaR levels are related in extreme situations. In clear, for institutions with high BC, the deltaCoVaR value is the smallest, which means the largest contribution to systemic risk, for institutions with BC at zero level, the deltaCoVaR is the highest (i.e. the lowest contribution to systemic risk). So AXA and Allianz contribute to systemic risk to a much greater extent than Phoenix. Figure 6 shows how does the deltaCoVaR depend on the mean BC. It can be seen that with the increase of BC of insurer vertices, the deltaCoVaR decreases (the contribution to systemic risk increases). Figure 7 presents a summary of MST and a diagram of the deltaCoVaR dependence on BC in the period when all analysed companies have the smallest deltaCoVaR, which happens to coincide with the middle of the crisis, that is October 17th, 2008. The tree during this period is such that AXA, Allianz and Aegon, having the largest BC and the largest contribution to systemic risk are directly related to each other. In Figure 8, we compile the contribution of AXA and Phoenix to system risk, i.e. the two companies that have the lowest and highest average deltaCoVaR, respectively. We observe the large differences between these companies in the determined market states: the normal one and two crisis states.

Fig. 4. Average value of BC in the period under consideration for individual insurance institutions

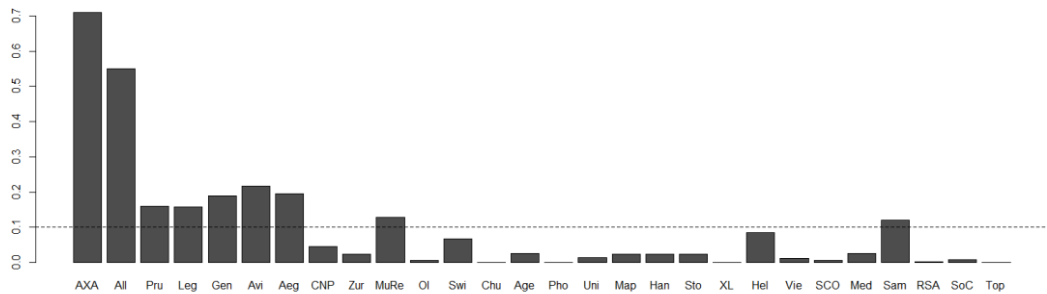

Source: Own study

Fig. 5 The average deltaCoVaR value over the period under consideration for individual insurers.

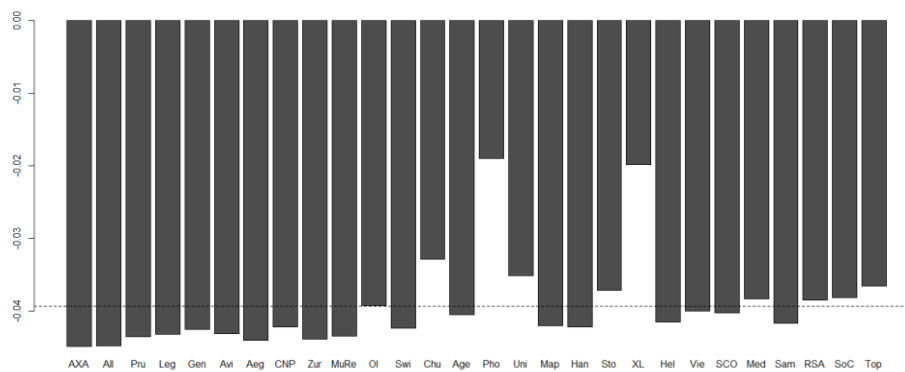

Source: Own study

**A DYNAMIC MST- deltaCoVar MODEL OF SYSTEMIC RISK IN THE EUROPEAN INSURANCE SECTOR.**

Fig. 6 The relationship between the average BC and deltaCoVaR values over the period under consideration for individual insurers

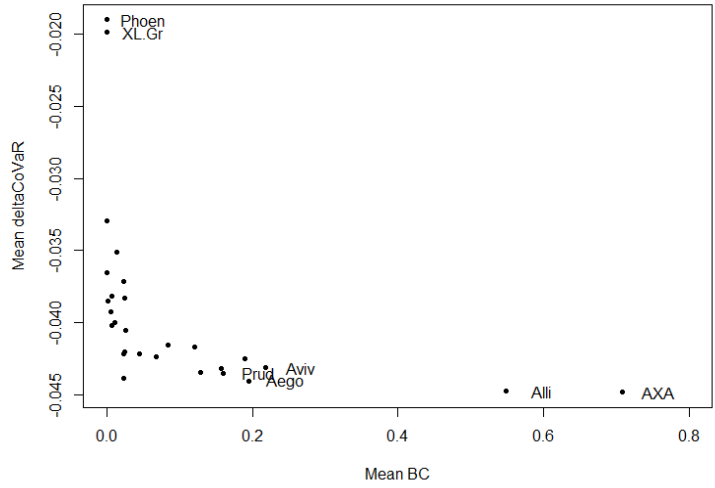

Source: Own study

Fig. 7 MST and relationship between BC and deltaCoVaR in the period with the lowest deltaCoVaR for AXA and Allianz

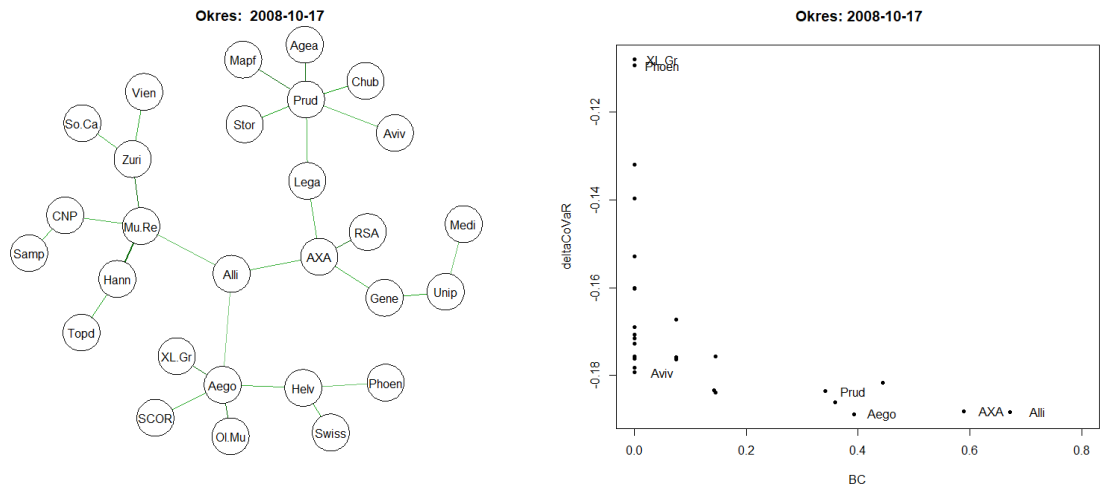

Source: Own study

Fig. 8 DeltaCoVaR for AXA and Phoenix in the period under consideration (left panel) and its distribution in the normal state of the market (N), during the subprime mortgage crisis (SMC) and the public debt crisis (PDC) (right panel)

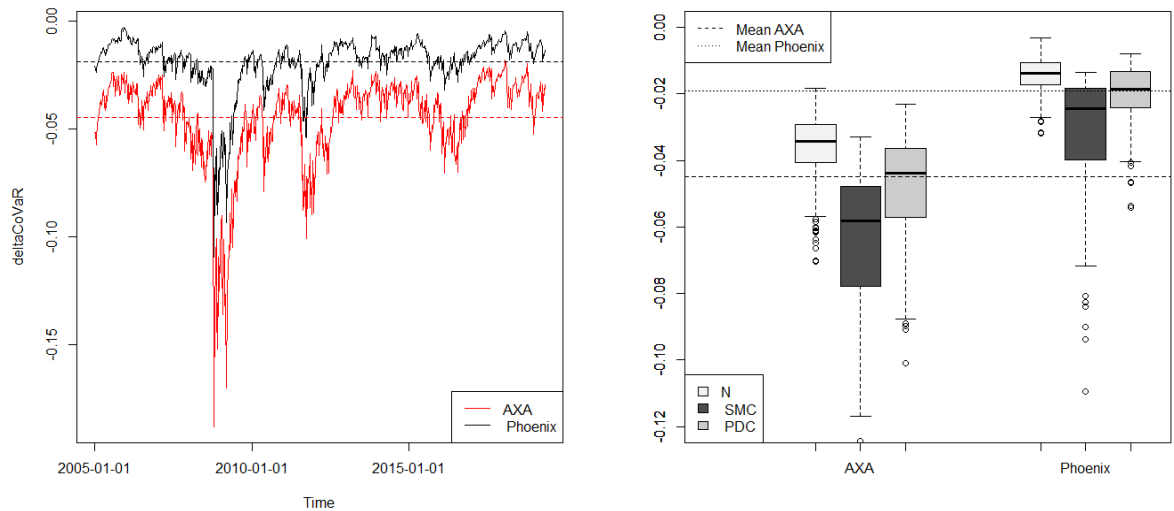

Source: Own study

## Conclusions

Analyzing the time series of topological APL, MD, BC indicators and the MST alpha indicator and the MST structure [citing work with Arxiva], we conclude that when the market is in a normal state, the series show volatility, but they do not have large amplitude. MSTs change their structure and the linkages between the companies is different depending on the time. We also note that to assess the potential risk of a rapid spread of the crisis, it is necessary to analyse all the four indicators. Indeed, with high correlation, the tree becomes very dispersed, which means that assessing the linkages only on the basis of the correlation coefficient could lead to erroneous conclusions. From the MST analysis, a clear shrinkage of the network can be seen in the periods of June 2nd, 2006 - August 17th, 2007, i.e. just before the subprime crisis and during its first phase, and in the period December 5th, 2008 - September 17th, 2010, i.e. before and at the beginning of the European public debt crisis. If MSTs are shrunk, it promotes potential propagation of financial problems. However, during the subprime crisis itself, the trees changed their appearance. They were relaxed: APL increased, MaxDegree decreased.

In this study, we analysed the 28 largest insurance companies from the point of view of the contribution of each institution to systemic risk in accordance with the currently used deltaCoVaR measure. The analysis of time series shows that for each of the companies in the period from 2005 to 2019 there is an obvious relation between its contribution to systemic risk and the structure of the network of connections (MST). The contribution of each company during the entire period remains at the same level, save for the clearly apparent period during which the deltaCoVaR delta decreases and, consequently, the contribution to the systemic risk increases, and this happens at the very center of the subprime crisis, October 17th, 2008. As the deltaCoVaR changes, the APL ratio increases. We should emphasize that for the entire analysed period, it reaches its maximum exactly on December 5th, 2008.

**A DYNAMIC MST- deltaCoVar MODEL OF SYSTEMIC RISK IN THE EUROPEAN INSURANCE SECTOR.**

The identified relationship between the contribution to systemic risk and the minimum spanning tree structure described by topological network indicators can be used in the construction of models whose task is to predict the possibility of systemic risk. The construction of this type of predictive models is the subject of further research by the authors.